\documentclass[%
reprint,
amsmath,amssymb,
aps,
prb,
]{revtex4-2}

\usepackage{graphicx}
\usepackage{dcolumn}
\usepackage{bm}

\begin{document}
	
	
	\title{Solving differential equations with Deep Learning: a beginner's guide}
	
	\author{Luis Medrano Navarro}%
	\affiliation{%
		Instituto de Nanociencia y Materiales de Aragón (INMA),CSIC-Universidad de Zaragoza,\\
		50009 Zaragoza, Spain
	}%

	\author{Luis Martin Moreno}%
	
	\affiliation{%
		Departamento de F\'isica de la Materia Condensada, Universidad de Zaragoza, Zaragoza 50009, Spain
	}%
	\affiliation{%
		Instituto de Nanociencia y Materiales de Aragón (INMA),CSIC-Universidad de Zaragoza,\\
		50009 Zaragoza, Spain
	}%

	\author{Sergio G. Rodrigo}%
	\email{sergut@unizar.es}
	\affiliation{%
		Departamento de F\'isica Aplicada, Facultad de Ciencias, Universidad de Zaragoza, 50009 Zaragoza, Spain
	}%
	\affiliation{%
		Instituto de Nanociencia y Materiales de Aragón (INMA),CSIC-Universidad de Zaragoza,\\
		50009 Zaragoza, Spain
	}%

\begin{abstract}
The research in Artificial Intelligence methods with potential applications in science has become an essential task in the scientific community last years.  Physics Informed Neural Networks (PINNs) is one of this methods and represent a contemporary technique that is based on the fundamentals of neural networks to solve differential equations. These kind of networks have the potential to improve or complement classical numerical methods in computational physics, making them an exciting area of study. In this paper, we introduce PINNs at an elementary level, mainly oriented to physics education so making them suitable for educational purposes at both undergraduate and graduate levels. PINNs can be used to create virtual simulations and educational tools that aid in understating complex physical concepts and processes where differential equations are involved. By combining the power of neural networks with physics principles, PINNs can provide an interactive and engaging learning experience that can improve students'  understanding and retention of physics concepts in higher education.
\end{abstract}

\maketitle

\section{Introduction}
Artificial Intelligence (AI) has brought about a major shift in how we solve problems, especially those computers previously struggled with or could not solve efficiently. For example, classifying handwritten digits was once a difficult task for computers, but now it can be easily done using Neural Networks (NN).


Science follows the trail of the significant progress made in AI last years, embracing these advancements and introducing novel techniques that could pave the way for future breakthroughs~\cite{app}.

Historically, one of the fundamental challenges in mathematics has been finding adequate solutions to differential equations. They are essential to our comprehension of the natural world and are extensively employed in numerous knowledge domains.

The initial attempts of AI using NNs for solving differential equations can be traced back to the 1990s~\cite{1998}. The method starts by choosing a trial function that satisfies the boundary conditions of the problem strictly by definition. Different problems thus require using different trial functions. Moreover, the method's validity depends on our ability to provide trial functions capable of approximating the solution, which often requires a good dose of imagination.  This presents a significant limitation for the method to be generalized to complex differential equations. 
More recently, new techniques have been proposed to overcome the previously cited difficulties.  Among these, we highlight here the Fourier Neural Operator Networks~\cite{ZhangProccML19} method, Deep Operator Networks (DeepONet)~\cite{LuNatMach21}, and Physics-Informed Neural Networks (PINNs).  The above methods use Deep Learning (DL) techniques to implement complex and intricate networks in different scenarios.  Fourier Neural Operator Networks combine the expressiveness of NNs with the mathematical structure of the Fourier series. The Approximation Theorem has inspired DeepONet, which suggests using deep networks in learning continuous operators from data. PINNs can be regarded as standard NNs, with the novelty that the loss function incorporates both the differential equation and the boundary/initial conditions. PINNs are already being used to solve differential equations in many fields, such as fluids physics~\cite{fluidos2}, quantum mechanics~\cite{cuantica3}, or photonics ~\cite{fotonica2}.   Compared to other methods, one of the advantages of PINNs is that they are relatively straightforward to grasp, requiring only a basic understanding of NNs and elementary calculus. 

In this work, we provide a beginner's guide on PINNs, introducing their fundamentals in an accessible manner.  For this, we show how they can be applied to a few examples of well-known Ordinary Differential Equations (ODEs), which are representative of equations appearing in physics. Specifically, we have selected three different ODEs that exhibit exponential decay solutions, harmonic oscillations or support solitary waves (solitons). 

We provide also the corresponding Python codes (supplied as Jupyter notebooks), which are fully described in the Supplementary Material (SM), all located in the following Github repository~\cite{github}. These notebooks can be used to reproduce all the results presented in this work and serve the reader to explore and find the solutions to other ODEs.

\section{Basics of Neural Networks}
Before explaining in depth how PINNs work, it is necessary to understand the basics of  NNs~\cite{nielsen}. NN mimics what we know about human neurons and brain tissue. In a human neuron, dendrites collect the input signals, which are then added inside the neuron, weighting them according to their importance. If this added value exceeds a certain threshold, the neuron triggers and sends a signal through the axon to other neurons. 

In analogy,  a mathematical neuron is defined as a function that takes $n$ real values as input and maps them into a real number through the composition of a linear transformation with the action of a nonlinear function. This is, given an input vector $\vec x=\{x_1,x_2,...,x_n\}$, the neuron returns $a(z)$, where $z=\sum_{i=1}^n \omega_i x_i - b$ is a weighted sum, and $a$ is the so-called \textit{activation} function. The coefficients $W=\{ \omega_i \}$ and $b$  are called \textit{weights} and \textit{bias}, respectively. 
Schematically, the action of a neuron can be expressed as follows: 
\begin{equation*}
    \begin{array}{c@{\quad \longrightarrow \quad}c@{\quad \longrightarrow \quad}c}
        \mathbb{R}^n &  \mathbb{R} & \mathbb{R}\\
        \{x_n\} & z = \sum_{i=1}^n \omega_i x_i - b & a(z)
    \end{array}
\end{equation*}

The threshold is encoded in the linear transformation for a mathematical neuron, that is, $\sum_{i=1}^n \omega_i x_i > b$. The activation function is chosen to mimic the triggering response of a real neuron. While the activation function in the first neuron reported (called \textit{perceptrons}~\cite{perceptron}) was chosen to be the Heaviside function (with values that changed abruptly from 0 to 1), the activation functions used nowadays are smooth as, in this case, the machinery of differential calculus has proven extremely helpful in the learning process.  
Currently, no definitive recipe tells us which is the best activation function for a particular problem. Still, the experience accumulated in the last few years provides some guidelines.  A smart selection of the activation functions is one of the crucial aspects when developing effective and efficient NNs.


\subsection{Neural Networks}
\begin{figure}[h]
  \centering
  \includegraphics[width=0.41\textwidth]{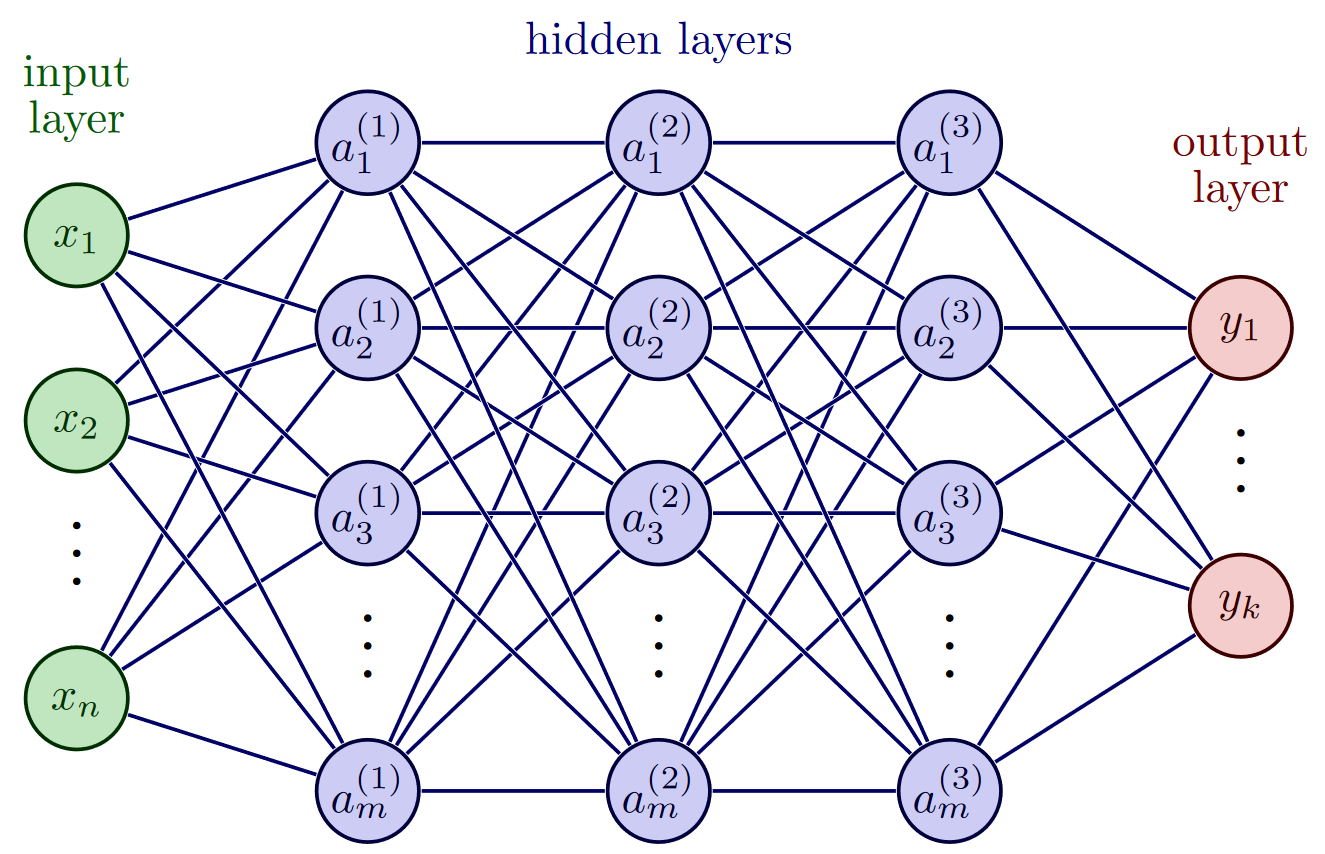}
  \caption{Schematics of a \textit{Deep} NN. Each instance $\vec x=\{x_1,x_2,...,x_n\}$ of the training data enters the input layer. These values are transformed through the neurons of the hidden layers by consecutive linear and nonlinear operations. Finally, the output layer returns the predictions of the NN, $\vec{y}_{NN}=(y_1,...,y_k)$.} 
  \label{red}
\end{figure}

A basic NN typically comprises a series of layers of interconnected neurons (see Fig.~\ref{red}). Information flows through the network from left to right. The first layer is the input layer $\vec x$ followed by the so-called hidden layers.  The output layer will provide the predictions of the NN and it will be denoted as $y_{NN}(\vec{x}, \theta)$, being $\theta = [W^k, b^k]$ for  $0\leq k\leq l$, and $l$ the number of layers. Therefore $\theta$ represents the set of weights and biases of all hidden layers at a particular training step and depends on the network's specific state during training.

The word \textit{deep} in Deep Learning is used to designate the case when there is a large number of hidden layers in the network. Curiously by today's standards, where Deep NN may have tens of hidden layers, even NNs with two hidden layers were already considered deep in the early days of AI research in this field. 

All the parameters we can manually change in a NN, such as the number of layers, the number of neurons per layer, the number of training epochs, etc., are called \textit{hyperparameters}. We must constantly play with these values for our NN to work correctly. The problem is that there are no clear rules for fine-tuning, so it has to be done manually, possibly with heuristic rules. 



\subsection{Loss function}
In standard uses of NNs, we have access to ``true'' data, i.e., many instances of pairs $(\vec x,\vec y_{true})$ for learning are available. This is referred to as supervised learning in the literature. As already explained, we expect the NN learns a transformation between the inputs and the outputs so the ``distance'' between the predictions and the actual values will ideally reach near zero values. 

This distance, between actual values and predictions by the NN, is codified in the so-called \textit{loss} function, $L$. The loss function depends on the weights and biases of all the neurons in the NN. There are infinite possible choices for the loss function, but some are standard in AI. The simplest one is to consider the Mean Squared Error (MSE) between the network's outputs/predictions $\vec y_{NN}(\vec x)$ and the true values $\vec y_{true}(\vec x)$ for all training data: 
\begin{equation} \label{ec:mse}
    MSE = \frac{1}{n}\sum_i |\vec y_{true}(\vec x_i)-\vec y_{NN}(\vec x_i)|^2
\end{equation}
, being $n$ the number of data points used to calculate this function. 

\subsection{The learning process}
Since the weights and biases used to be initialized with random values, the initial predictions $\vec y_{NN}$ are not related to $\vec y_{true}$, and the initial value of the loss function is high. Therefore, the learning process, referred to as training, minimizes the loss function by gradually modifying the weights and biases. The training is usually divided into epochs.  An epoch refers to one complete iteration over all the training data. The number of epochs is the number of times the NN sees and learns from a training dataset. 

Usually, the minimization of the loss function is based on the gradient descent method. In each gradient descent step, we update the values of each weight $\omega_i$ and each bias $b_i$ according to the following equations (the gradients)
\begin{equation}\label{ec:descenso_pesos}
    \omega_{i}^{n+1} = w_{i}^n - \eta\frac{\partial L}{\partial \omega_i}
    \hspace{1cm}
    b_{i}^{n+1} = b_{i}^n - \eta\frac{\partial L}{\partial b_i}
\end{equation}
where the parameter $\eta$ is known as the \textit{learning rate}.  As the gradient points towards the direction of the maximum increase of a function,  the search for a global minimum is done by moving in the opposite direction of the gradient.

The learning rate is another hyperparameter to tune while searching for an optimum NN model. The overall training process is like descending a mountain to reach its lowest point.  Taking too large steps (large 
$\eta$) would make it hard to find the lowest point on the valley because when we get close and take the next step, we might miss it and end up slightly back up the mountain. To improve the accuracy of the NN predictions, we can reduce this parameter at the cost of increasing the execution time of the algorithm. 

The learning process is prone to several challenges, such as the algorithm becoming trapped in local minima instead of discovering the global minimum.  Selecting an appropriate optimizer can address these and other issues. There are numerous optimizers available for neural networks. Stochastic Gradient Descent (SGD) consist in an improved version of the gradient descent algorithm. It updates the model parameters in small batches, randomly selected from the whole training dataset. SGD is an effective optimizer that is widely used due to its simplicity and ease of implementation. Another optimizers include popular ones such as Adam and RMSprop ~\cite{geron}. Adam is particularly good when dealing with noisy data, and it is computationally efficient because it implements adaptive learning rates. On the other hand, RMSprop adapts learning rates for each parameter based on the average of its recent magnitudes, which allows it to converge faster and more reliably than other methods in some cases. 

\subsection{Backpropagation}
Gradient descent allows us to minimize the loss function but does not tell us how to calculate the gradients. We could calculate the partial derivatives numerically:
\begin{equation}
     \lim_{\Delta x_i \rightarrow 0} \frac{y(x_1,...,x_i+\Delta x_i,...,x_n) - y(x_1,...,x_i,...,x_n)}{\Delta x_i}
\end{equation}
However, this process requires a vast number of operations and would make learning unbearably slow and inefficient.

One extremely efficient way to compute the gradient is using Automatic Differentiation (AD). There are several techniques available with different advantages and disadvantages.

In the context of NN the method to calculate the gradients is Backpropagation, a form of reverse-mode AD that uses the chain rule to compute gradients efficiently~\cite{backpropagation}. This algorithm enabled the first major AI revolution. It is a two-step process.  First, a forward pass generates a set of activations of all neurons. Then, all the partial derivatives necessary to update the weights and bias are obtained using the chain rule~\cite{nielsen}. The Backpropagation algorithm relies on two main assumptions regarding the loss function $L$. The first assumption is that the loss function can be expressed as an average over the loss functions for individual training examples. The second assumption is that the loss function can be expressed as a function of the outputs generated by the neural network. As an example, the  MSE loss function described by Eq.~\ref{ec:mse} holds all the requirements. 

\begin{figure*}
  \includegraphics[width=1.0\textwidth]{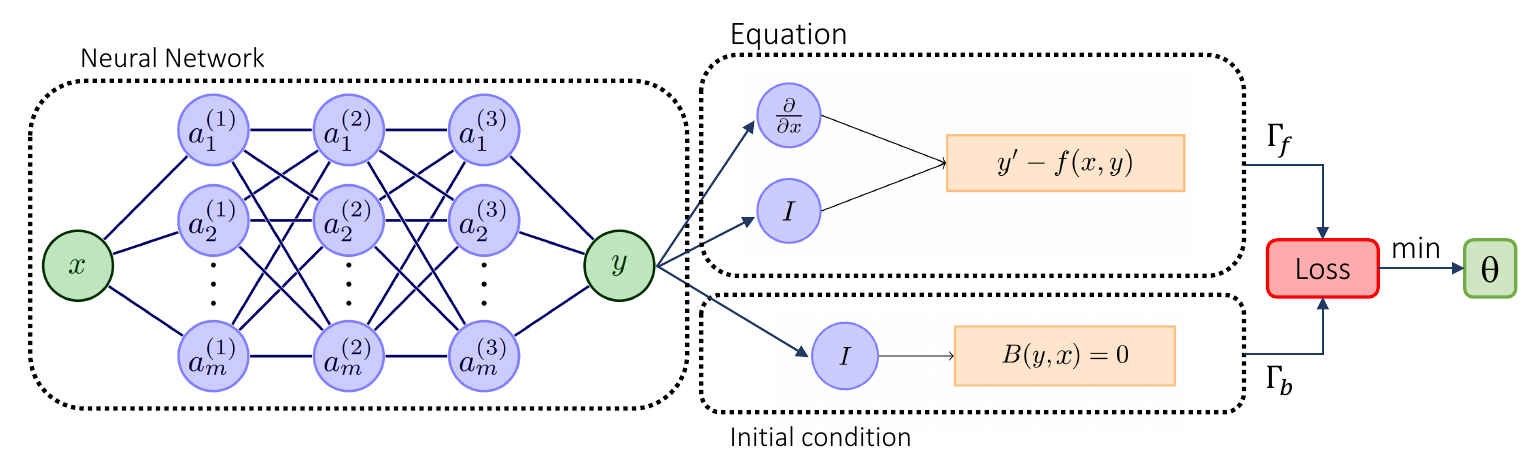}
  \caption{PINN workflow for first-order ODEs. The approach involves building a basic NN, consisting of one input neuron for $x$ (points in the region where we search the solution) and one output neuron for $y_{NN}(x)$, which is obtained by performing a forward pass through the NN. After that, we can obtain the derivative $y'=dy/dx$ thanks to Automatic Differentiation (essential part of the backpropagation algorithm).  Then the losses associated with the ODE and the initial condition are computed. Finally, the optimizer updates the weights and biases. This process is repeated with multiple input points and through many training epochs.}
  \label{PINN}
\end{figure*}

\section{Physics Informed Neural Networks}\label{pinn}
This section demonstrates that NNs comprise all the elements necessary to solve a differential equation. PINNs are regular NNs that incorporate both the differential equation and its boundary/initial conditions within the loss function. This way,  PINNs can be applied to a broad type of differential equations, whether ordinary or partial, with one or several variables, single equations, and even systems of equations. It should be noted that, despite its name, PINNs are actually mathematical solvers for Partial Differential Equations (PDEs) and do not inherently rely on any physics-specific information. 

Let us start by providing a general definition of PINNs and how they apply to solve PDEs.  Typically, a PDE can be expressed as follows:
\begin{equation} \label{eq:diff.equ.gen}
        D\left( \vec{x},y(\vec{x}),\frac{\partial y}{\partial x_1},...,\frac{\partial y}{\partial x_d},\frac{\partial^2 y}{\partial x_1\partial x_1},...,\frac{\partial^2 y}{\partial x_1\partial x_d},... \right)=0 
\end{equation}
 where $\vec{x}=(x_1,...,x_d)$ is a $d$-dimensional vector defined in a region $\Omega \subset \mathbb{R}^d$, and $y(\vec{x})$ is the solution Eq.~\ref{eq:diff.equ.gen}, satisfying the boundary conditions given by:
\begin{equation}
        B(\vec{x},y(\vec{x}))=0 \hspace{0.3cm} \vec{x} \in \partial\Omega ,
\end{equation}
where $\partial\Omega$ represents the boundary of $\Omega$.

What enables a NN to find the solution of a given PDE is its ability to represent all the mathematical objects present in Eq.\ref{eq:diff.equ.gen}: $\vec{x}$, $y(\vec{x})$ and the partial derivatives of $y(\vec{x})$. The input layer of the network corresponds to $\vec{x}$ ($d$ input neurons), and the output of the last layer represents the solution $y(\vec{x})$ (only one neuron) for the PDE.  Largely, this is due to the universality theorem of NNs\cite{universal}, which states that they can represent any function, and so it does so with the solution of a particular PDE. But it also depends on the capacity to find partial derivatives of $y(\vec{x})$, through AD, the NN can provide all the partial derivatives necessary to solve the differential equation on each training step.

To solve the differential equation, we choose a set of points $N_{B}$ at the boundary $\partial \Omega$ (denoted as $\Gamma_B$) and a set of $N_{D}$ points in the interior of  $\Omega$ (denoted as $\Gamma_D$). The set $\Gamma$ is defined by the union of these two sets, $\Gamma =  \Gamma_B \cup \Gamma_D =\{\vec x_0, \vec x_1,...,\vec x_M\}$, with $M =N_{B} +N_{D}$. The set  $\Gamma$ contains the points where $y_{NN}$ is evaluated in the training process.

To train the PINN, we also need a loss function, which is defined as a weighted sum of two terms:
\begin{equation}\label{ec:loss}
        \mathbf{L}(\theta, \Gamma)=\omega_D\mathbf{L}_D(\theta, \Gamma_D) + \omega_B\mathbf{L}_B(\theta, \Gamma_B).
\end{equation}
The term $ \mathbf{L}_D(\theta, \Gamma_D) $ should measure how well the PDE is being satisfied by $y_{NN}$ at points in the set $\Gamma_{D}$, while $ \mathbf{L}_B(\theta, \Gamma_B)$ should measure how well the boundary condition is satisfied by $y_{NN}$ at points in the set $\Gamma_{B}$.

One common practice that we are going to follow in this work is to choose the following:
\begin{eqnarray}
        \mathbf{L}_D(\theta, \Gamma_D) = \\ \frac{1}{N_D}\sum_{\vec{x}\in\Gamma_D}\left| D\left( \vec{x},y_{NN}(\vec{x}),\frac{\partial y_{NN}}{\partial x_1},...,\frac{\partial^2 y_{NN}}{\partial x_1\partial x_1},... \right) \right|^2
\end{eqnarray}
and 
\begin{equation}
        \mathbf{L}_B(\theta, \Gamma_B) = \frac{1}{N_B}\sum_{\vec{x}\in\Gamma_B}\left| B(y_{NN},\vec{x}) \right|^2 .
\end{equation}
Finally, in Eq.\ref{ec:loss}, $\omega_D$ and $ \omega_B$ are hyperparameters that must be tuned. These are usually set to one, as we will do in the examples that follows.

As with other uses of NN, the loss function is minimized iteratively. If its minimum value is near zero, the solution $y_{NN}$ approximately fulfills the differential equation and the boundary condition at the chosen points in $\Gamma$. The general validity of the solution can be estimated by computing the value of the loss functions at points $\vec{x}$ not used in the training process. 

\section{First order Ordinary Differential Equations}
In the previous section, we described the general formalism of PINNs. Let's now apply it to the simpler case of a 1st-order Ordinary Differential Equation (ODE), which can be written in its general form as:
\begin{equation}\label{ec:firstODE}
    \left\{
        \begin{array}{l}
            y'(x)=f(x,y)\\
            y(x_0)=y_0
        \end{array} 
        \right.\qquad
\end{equation}
where $y'(x)=\frac{dy}{dx}$, being $f(x,y)$ an arbitrary function of $x$ and $y(x)$. The initial condition is the value of the function $y$ at $x_0$. 

Figure~\ref{PINN} shows the workflow of a PINN adapted to 1st-order ODEs. By the notation used in the preceding section, 
\begin{eqnarray}
    D(x, y, y')=0 \rightarrow y'(x) -f(x,y)=0 \\
    B(y_0,x_0)=0 \rightarrow  y(x_0)-y_0=0
\end{eqnarray}
The ODE loss is then:
\begin{equation} \label{ec:Lf}
    \mathbf{L}_D(\theta, \Gamma_D) = \frac{1}{N_D}\sum_{i=1}^{N_D} \left [ y'_{NN}(x_i,\theta)-f(x_i,y_{NN}(x_i,\theta)\right]^2
\end{equation}
The initial condition loss is written as:
\begin{equation} \label{ec:Lb}
    \mathbf{L}_B(\theta, \Gamma_B) = \left| y_{NN}(x_0,\theta)-y_0 \right|^2 
\end{equation}
A successful training ends up with a set of optimum weights and biases values $\theta_{opt}$, so that $\mathbf{L}_B(\theta_{opt}, \Gamma_B) \approx 0$ and $\mathbf{L}_D(\theta_{opt}, \Gamma_D) \approx 0$.

\begin{figure*}
  \centering  
  \includegraphics[width=1.0\textwidth]{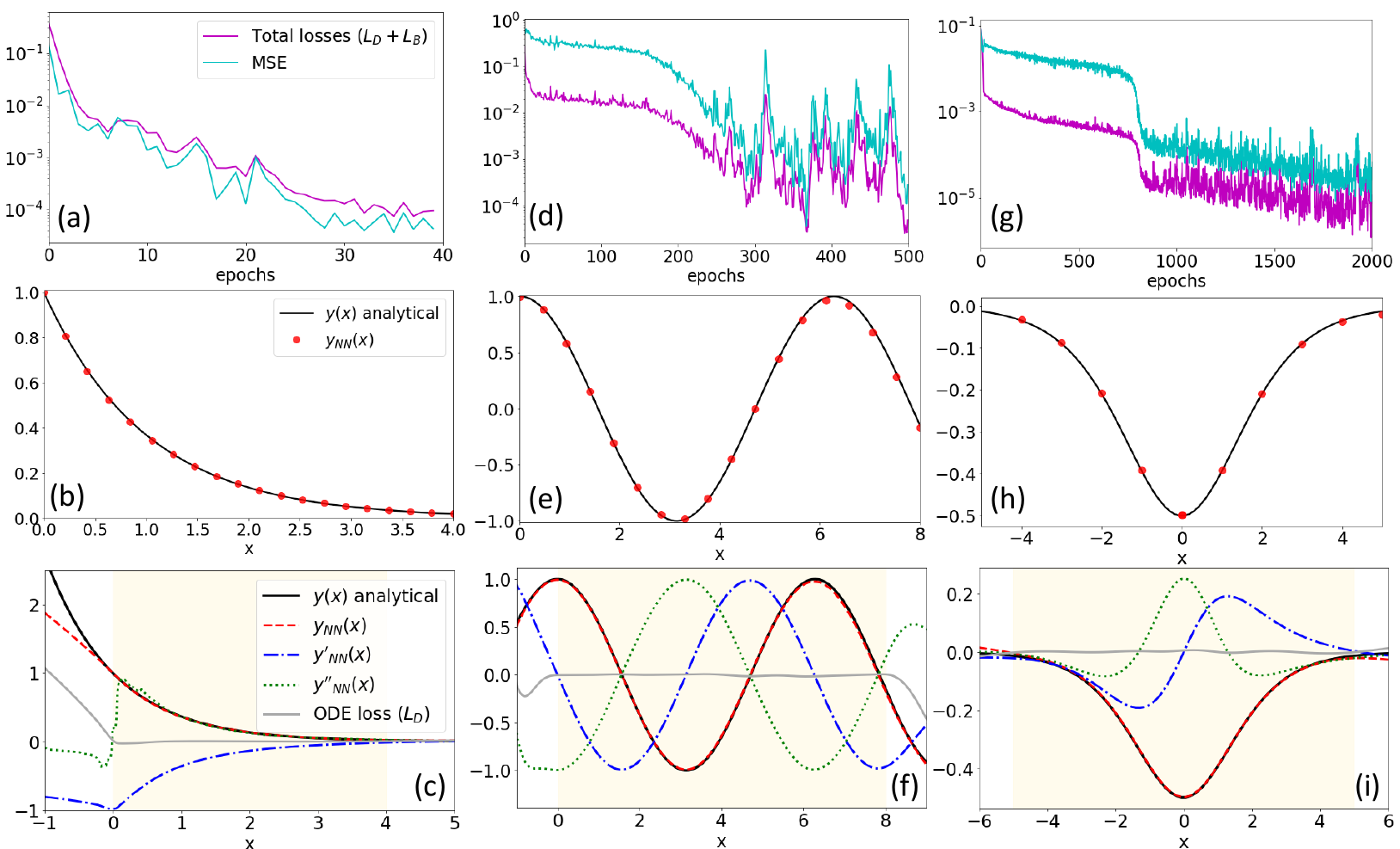}
  \caption{Three representative physical examples described by ODEs are studied using PINNs. Panels [a-c] present the analysis of a 1st-order ODE, corresponding to the exponential decay solution (Eq.~\ref{ec:exp}); [d-f] shows the study of the harmonic oscillator as a representative example of a 2nd-order ODE (Eq.~\ref{ec:osc}); and [g-i] is devoted to the Korteweg-de Vries 2nd-order nonlinear ODE, whose solution is a soliton (Eq.~\ref{ec:gp}).  The first row displays the training evolution as a function of the number of epochs. Two indicators of the performance of the NN are plotted: the total loss ($L_D$ + $L_B$) and the MSE. The last measures the distance per point between $y_{analytic}$ and $y_{NN}$, evaluated at the training points. The second row compares the analytical solutions (solid lines) and those predicted by the PINNs at the training coordinates (dots). The third row displays the analytical solutions and the predictions for $y(x)$ and its derivatives as a function of $x$. The results of this row were obtained from previously unseen points inside and outside the training interval (shadowed regions), which were utilized as validation data for the PINN. The ODE loss ($L_D$) is also included (solid gray line) as a reference.     
  }
  \label{results}
\end{figure*}

The first example we describe is a well-known 1st-order ODE, where the derivative of a function is proportional to minus the value of the function (in what follows, we choose the units such that both the proportionality constant and $y(0)$ are the unity):
\begin{equation}\label{ec:exp}
    \left\{
        \begin{array}{l}
            y'(x)+y(x)=0 \hspace{0.3cm}\\
            y(0)=1
        \end{array} 
        \right.\qquad
\end{equation}
, whose exact solution is  $y(x)=\exp(-x)$.

This type of ODE equation describes, for instance, radioactive decay, how hot objects cool down, and the time dependence of the charge on a capacitor in a resistor-capacitor electrical circuit. But it also appears in problems beyond physics. For example, in chemistry, the rate at which a chemical reaction occurs is often proportional to the concentration of the reactants. In sociology, the rate of change of a population is sometimes proportional to the size of the population. 

The main results are presented in Fig.~\ref{results} (Section~2 in the SM includes the corresponding Python implementation and its description). 

It should be noted that the results may vary across different runs of the codes due to the statistical behavior of certain parts of the algorithm. In the example of this section, we achieved a good performance with a NN that had just 3 hidden layers with 50 neurons each and chose 20 points in the training interval ($0 \leqslant x \leqslant 4$). We utilized the \textit{Adam} optimizer with a learning rate of 0.001. The hyperparameters measuring the relative relevance of ODE and initial condition losses are chosen as $\omega_D=\omega_B=1$ (see Section~2 in the SM, for further details).
 
In this case, total loss values  ($\mathbf{L}_D+\mathbf{L}_B$) around $1.0\cdot 10^{-4}$ provided correct NN predictions. A total of 100 simulations were conducted setting this total loss as a threshold. The result is that, on average, it takes approximately 42 epochs to achieve this threshold. The standard deviation is also very low (9 epochs). 

The total loss as a function of the epochs is shown (logarithmic scale) in Fig.~\ref{results}(a) with the magenta line, for one of the simulations conducted. Overall, the total loss decreases with the number of epochs. In the same panel, the MSE is measuring the distance between the exact analytical solution and the calculated by the PINN (see Eq.~\ref{ec:mse}).  This metric is used to estimate the accuracy of the predicted solution, and it has not been used in the training process (PINNs make not use of external data to work), as it requires knowledge of the analytical solution beforehand. 

Both metrics in Fig.~\ref{results}(a) become negligible after only 40 epochs of training. This indicates that the PINN provides a good approximation for the solution of the ODE on the training points, which is confirmed in Fig.~\ref{results}(b), where the solid line corresponds with the exact solution, while circular symbols provide the values of $y(x)$ predicted by PINN.

Although the NN may make accurate predictions for the training points, this does not necessarily imply that it will generalize well to new, unseen data points. A crucial consideration is thus the ability of the PINN to interpolate and extrapolate results accurately. We use a validation set, with many points chosen within and outside the training interval. The results are summarized in Fig.~\ref{results}(c). The highlighted region marks the training interval. We see how the predictions of the NN (dashed red line) nicely interpolate within the training interval at points not seen by the NN during training.   However the PINN is less reliable for extrapolation; the prediction deteriorates quickly away from the interval used in training.

To better understand how the PINN generalizes, we can take a look at the predictions of ${y(x), y'(x),y''(x)}$  and the ODE loss ($L_D$) within the validation interval. These quantities are shown in Figure~\ref{results}(c). The ODE loss vanishes when the NN prediction is correct, indicating an excellent approximation of the solution. This is the only possible indicator of the quality of the solution outside the training points in case the analytic or numerical solution is unknown. From a pedagogical point of view is interesting to observe how the NN is able to obtain the correct derivatives of the function. The first derivative looks pretty good inside the training interval. However, $y''(x)$ does not exactly coincides what it would be expected within the range $0<x<1$, although it is within the training interval. This could be the cause of the deterioration observed for predictions outside the training interval.

PINNs offer several benefits compared to traditional numerical methods (e.g., finite element methods). One of the most significant advantages is that they eliminate the need to create a mesh for the entire domain of the differential equation, which is a common challenge in classical algorithms. In our example, only a few points were necessary for a correct solution. 

\section{Second order Ordinary Differential Equations}
The description of physical systems frequently relies on 2nd-order ODE, like in many models of Newton's mechanics. 

In this section, as an example, we will find the solutions for the harmonic oscillator using PINNs.  Examples of physical systems described by a harmonic oscillator are a mass attached to a spring,  a  pendulum consisting of a mass hanging from a string swinging back and forth with small amplitude, and an electrical circuit composed of an inductance and a capacitance in series.
To simplify the description, we will use units and initial conditions such that the ODE is described as:
\begin{equation}\label{ec:osc}
    \left\{
        \begin{array}{l}
            y''(x) + y(x)=0 \hspace{0.5cm}\\
            y(0)=1, y'(0)=0
        \end{array} 
        \right.\qquad
\end{equation}
,whose analytic solution is $y(x)=\cos(x)$. The reader will have noticed that we do not use the usual notation for the harmonic oscillator solution, which usually describes de dependence of some quantity with time $t$. To maintain standard AI terminology and consistent notation throughout the paper, we denote that quantity as $y$ and ``time'' $t$ with $x$. 

We solve Eq.\ref{ec:osc} using the PINN machinery described before (the interested reader can find in Section~3 of the SM the Python implementation). After slightly tuning the hyperparameters, we get total losses and MSE values as small as the ones for 1st-order ODE studied in the previous section (Fig.~\ref{results}(d)), for an example of this PINN evolution. As in the previous case, we imposed a total loss threshold ($1.0\cdot 10^{-4}$) that produces pretty exact NN predictions with this PINN. Across 100 simulations, the NN reached this threshold after an average of 326 epochs of training, with a standard deviation of 36 epochs.

We can observe, however, a distinct behavior, the appearance of strong oscillations. This can be due to the stochastic behavior of the optimizer along the training process so that the model can be trapped momentarily in a local minimum of the loss function. If the loss does not improve after several epochs, the Adam optimizer can change the learning rate, which may be the most likely cause of these jumps in the loss.  To minimize the effect of these fluctuations, during the training process, we keep saving the best model.

Figure~\ref{results}(e) displays the predictions of the PINN within the training interval of $0\leqslant x \leqslant 8$. Like the 1st-order ODE analysis, the minimal total losses and MSE indicate an excellent agreement between the analytical solution and $y_{NN}$.

On the other hand, Figure~\ref{results}(f) summarizes the robustness of the PINN predictions within the training interval, with limited performance beyond it. When applied to validation points, the PINN provides perfect interpolation. Additionally, the PINN solution remains valid slightly outside the training region. As anticipated, the ODE loss within the training interval is nearly negligible.

Figure.~\ref{results}(f) also provides an excellent visualization of how the PINN predictions fulfill the mathematical relations between the $y(x)=\cos(x)$ function and its first two derivatives. 

\section{Second order nonlinear Ordinary Differential Equations}
We present in the last section the case study of a nonlinear ODE (refer to Section~4 in the SM for the Python implementation). Nonlinear physics explain many physical phenomena beyond linear PDEs' scope. For instance, under intense laser excitation, certain materials display nonlinear optical responses, resulting in the appearance of colors that were not initially present in the excitation beam. Solitons are another example of the effects caused by the nonlinearities in materials.  These waves can keep their shape and velocity, even in media where completely distorted waves would be expected in the absence of nonlinear effects.

One nonlinear ODE that appears in several physical situations is the Korteweg-de Vries (KdV) equation, which can model solitons in ocean waves, fiber optics modes, and Bose-Einstein condensates in quantum mechanics.

The general solution of the KdV equation has complex dependencies between time and space. Still, it can be constrained to maintain its shape. Mathematically, $\phi(X,t)=y(X-vt)=y(x)$, where $v$ is the solution speed.  With this constrain, the following 2nd-order ODE mathematically describes a soliton in the KdV model:
\begin{equation}\label{ec:gp}
    \left\{
        \begin{array}{l}
            y''(x) - y(x) - 3y^2(x)=0  \\            
            \hspace{1cm}y(0)=-1/2, y'(0)=0
        \end{array} 
        \right.\qquad
\end{equation}

Another advantage of using the KdV equation as an example is that its analytical solution is known:  $y(x) = - \frac{1}{2}\cdot sech ^2(\frac{x}{2})$. By changing the variable $x \rightarrow X-t$, the equation describes a soliton that moves to the right with a velocity of one (in the chosen set of units).

We analyze the total loss and MSE in Fig.~\ref{results}(g). The complexity of this equation, and probably the presence of the nonlinear term ($y^2(x)$), made it necessary to reduce the learning rate to 0.0001, i.e., by a factor of 10, as compared with the previous ODEs.  Correspondingly,  the number of epochs had to be increased to ensure the loss reached a sufficiently small value. Similar to the previous PINNs experiments, we conducted 100 simulations while enforcing a total loss threshold of $1.0\cdot 10^{-5}$. This value was determined empirically and has been observed to yield reliable NN predictions. On average, it takes approximately 1422 epochs to achieve this loss threshold, with a standard deviation of 623 epochs. It is worth noting that the distribution in this case is non-normal (see Section~5 in the SM) and a more insightful analysis requires the use of median and mode statistics, where the median represents the middle value of the dataset and the mode represents the most common value. In the 2nd-order nonlinear ODE case the median and mode were 1256 and 1225, respectively. This last explains why only 12\% of the simulations required 2000 epochs or more to converge to the chosen threshold.

The main result, that is, the solution of the soliton, is achieved in the interval $-5\leqslant x \leqslant 5$ (Fig.~\ref{results}(h)) using only 10 training points. Then, again, the predicted solution, its derivatives, and the ODE loss are calculated and shown in Fig.~\ref{results}(i). It is interesting to note how this PINN is able to find a exact solution with so sparse mesh of 10 points. 

\section{Conclusion}
We have presented an introductory guide to PINNs, a recently proposed method to solve differential equations with AI. From a pedagogical point of view, by combining the power of NN with the knowledge encoded in differential equations, PINNs offer a promising approach to teaching physics. By using PINNs, students can explore and experiment with virtual simulations of a large variety of physical systems, gaining a deeper understanding of the underlying concepts and equations. We have described several examples of commonplace differential equations in physics, intending to introduce this new technique to both undergraduate and graduate students and also teachers.  We provide as SM the codes we implemented and used in all the example, to facilitate the learning and teaching of this exciting method.  We hope that our easy-to-understand code for PINNs enable students and teachers to experiment with it.

\section{References}

\end{document}


\title[]{Supplementary material - Solving differential equations in physics with Deep Learning: a beginner’s guide}
\author{Luis Medrano Navarro}
\address{Instituto de Nanociencia y Materiales de Aragón (INMA), CSIC-Universidad de Zaragoza, 50009 Zaragoza, Spain}
\ead{780070@unizar.es; ORCID=0000-0003-3246-9383}
\author{Luis Martin Moreno}
\address{Instituto de Nanociencia y Materiales de Aragón (INMA), CSIC-Universidad de Zaragoza, 50009 Zaragoza, Spain}
\address{Departamento de Física de la Materia Condensada, Universidad de Zaragoza, Zaragoza 50009, Spain}

\ead{lmm@unizar.es; ORCID=0000-0001-9273-8165}
\author{Sergio G Rodrigo}
\address{Departamento de Física Aplicada, Facultad de Ciencias, Universidad de Zaragoza, 50009 Zaragoza, Spain}
\address{Instituto de Nanociencia y Materiales de Aragón (INMA), CSIC-Universidad de Zaragoza, 50009 Zaragoza, Spain}
\ead{sergut@unizar.es; ORCID=0000-0001-6575-168X}
\vspace{10pt}

%
\vspace{2pc}

%
\maketitle

\renewcommand{\contentsname}{}
\tableofcontents

\section{Deep learning PINNs with Tensorflow-Keras}
In the following, we describe the \textit{Python} codes that were implemented to conduct the calculations of the paper. These codes can be found in the following Github repository~\cite{github}. We provide a comprehensive explanation of them, such that it can be utilized effectively and modified according to one's requirements. The Supplementary Material of this paper also includes three \textit{Jupyter} notebooks, so the reader can practice with the examples provided and build its own PINNs.

To implement our PINNs, we have used Keras and Tensorflow libraries~\cite{chollet}. Tensorflow consists of a set of programming libraries to operate with tensors. With Tensorflow, it is possible to implement neural networks (NN) from scratch. However, being a low-level library, its learning and use are relatively complex. On the other hand, Keras is a high-level Application Programming Interface (API) where it is easier to create complex architectures with NN. We have used Keras as the backbone of the implementation, but it has been necessary to use Tensorflow to generate the Ordinary Differential Equation (ODE) loss functions described in the paper.  

\section{Example 1: 1st order Ordinary Differential Equations}
The first-order ODE to be solved with the use of PINNs is:
\begin{equation}\label{ec:exp}
    \left\{
        \begin{array}{l}
            y'(x)+y(x)=0 \hspace{0.3cm} with \hspace{0.3cm} 0 < x < 4 \\
            y(0)=1
        \end{array} 
        \right.\qquad
\end{equation}
, whose exact solution is  $y(x)=\exp(-x)$.

\subsection{Main libraries}
To begin with, we start the loading of essential packages for the algorithm. Primarily, we load Tensorflow, and in addition, we utilize Numpy for mathematical and array operations, and Matplotlib for generating plots. We further import from Keras two types of layers (\textit{Input} and \textit{Dense}) and the \textit{Adam} optimizer. 

\begin{python}
# Tensorflow Keras and the rest of the packages
import tensorflow as tf
from tensorflow.keras.layers import Input,Dense
from tensorflow.keras.optimizers import Adam
import numpy as np
import matplotlib.pyplot as plt
\end{python}

\subsection{Definition of the PINN} \label{defpinn1}
The next lines of code can be considered the core of the PINN algorithm. In there we create the loss function in Tensorflow using the differential equation information. 

The Python code defines a custom Keras model class \textit{ODE\_1st} that inherits from the \textit{tf.keras.Model} class. The \textit{train\_step} method of this class implements the training loop for the model (see Ref.~\cite{geron}, for an introductory description of custom objects using Tensorflow-Keras). Inside the \textit{train\_step} method, the \textit{data} argument is a tuple that contains the training inputs \textit{x} and the analytical (exact) solution \textit{y\_exact} at these input points. 

The word $self$ in this code means $model$. So, for example,  $self(x0,training=True)$ calculates the model's prediction at the point of the initial condition $x_0$. This is why the result is \textit{y0\_NN} in the corresponding line of code, the prediction of the NN at this point.

Let's describe the different parts of the Python code:
\begin{enumerate}
    \item
    The method starts by defining the initial conditions for the PINN, as \textit{x0} and \textit{y0\_exact}, which are set as constant tensors. It is mandatory that all variables defined throughout the code are in the Tensorflow format.
    \item 
    The code then computes the gradients of the output \textit{y\_NN} with respect to the input \textit{x} and the initial condition \textit{y0\_NN}, using two nested \textit{tf.GradientTape} contexts. The \textit{tf.GradientTape} is the part of Tensorflow dedicated to Automatic Differentiation (AD). This is a very efficient way of calculating derivatives using NN, allowing the NN to solve differential equations.  In particular, Tensorflow uses what is called reverse AD, based on the mathematical properties of dual numbers~\cite{geron}.  
    \item 
    The loss function is then calculated using the computed gradients, and the initial conditions are included in the loss calculation. The loss is a sum of two contributions of $self.compiled_{-}loss$, which in our case is the Mean Squared Error (MSE) function, described in the article. The line $self.compiled_{-}loss(dy_{-}dx_{-}{NN}, -y_{-}{NN})$ returns the error in the differential equation ($|y'+y|^2$). The code line calling to $self.compiled_{-}loss(y0_{-}{NN}, y0_{-}{exact})$ is the error in the initial condition ($|y(0) - 1|^2$).
    \item 
    Finally, the gradients of the loss function with respect to the trainable weights of the model are then computed using the top level \textit{tf.GradientTape}, and the \textit{Adam} optimizer is used to update the weights and biases based on these gradients. Finally, the model metrics (loss and MSE) are updated, and the method returns a dictionary of the updated metrics.
\end{enumerate}

\begin{python}
class ODE_1st(tf.keras.Model):                  
    def train_step(self, data):            
        # Training points 
        #and the analytical (exact) solution at these points    
        x, y_exact = data                
        # Initial conditions for the PINN        
        x0=tf.constant([0.0], dtype=tf.float32) 
        y0_exact=tf.constant([1.0], dtype=tf.float32) 
        # Calculate the gradients and update weights and bias
        with tf.GradientTape() as tape:   
            # Calculate the gradients dy/dx          
            with tf.GradientTape() as tape2:   
              tape2.watch(x0)
              tape2.watch(x)             
              y0_NN = self(x0, training=True)
              y_NN  = self(x, training=True)                 
            dy_dx_NN= tape2.gradient(y_NN,x)         
            #Loss= ODE+ boundary/initial conditions                              
            loss=self.compiled_loss(dy_dx_NN, -y_NN)\
                +self.compiled_loss(y0_NN,y0_exact) 
        gradients = tape.gradient(loss, self.trainable_weights)
        self.optimizer.apply_gradients(zip(gradients, self.trainable_weights))
        self.compiled_metrics.update_state(y_exact, y_NN)
        return {m.name: m.result() for m in self.metrics}
\end{python}

\subsection{Run the PINN}
The code of this section defines and trains the NN model to solve the differential equation using the PINN approach. Here's what the code does:
\begin{enumerate}
    \item 
    \textit{n\_train}, \textit{xmin}, and \textit{xmax} define the number of training points and the range of the input values.    
    \item
    \textit{x\_train} is a 1D NumPy array of size \textit{n\_train} containing equally spaced input values between \textit{xmin} and \textit{xmax}. $x0$ is the initial condition for the PINN, which is set to 0.0. The first element of \textit{x\_train} is replaced by $x0$.
    \item
    \textit{y\_train} is a 1D tensor of size \textit{n\_train} containing the true solution to the differential equation at the corresponding input values in \textit{x\_train}. The true solution is computed using the \textit{tf.exp} function in this example.
    \item
    The NN model is defined with an input layer, three hidden layers with 50 neurons each, and an output layer with a single output neuron. Note that the activation function of all neurons is $elu$, except for the last one, which does not apply any activation function (\textit{activation=None}).
    \item
    The model is compiled using: i) MSE as the loss function (here is ``written'' the ODE and its initial conditions through the custom Keras model class \textit{ODE\_1st}, as defined in the previous section); ii) the Adam optimizer with a learning rate of 0.001; and iii) the MSE metrics (distance between $y_{NN}$ and $y_{exact}$).
    \item
    The \textit{model.fit} method is used to train the model for 50 epochs, with a batch size of 1, using \textit{x\_train}  as inputs. Note that  \textit{y\_train} is used only to estimate the accuracy of the predicted solution, and it has not been used in the training process (PINNs make not use of external data to work). Therefore, this requires to know the analytical solution beforehand.
    \item
    Finally, the \textit{history} object returned by \textit{model.fit} contains information about the training progress, such as the total loss and metric values at each epoch.
\end{enumerate}

\begin{python}
n_train = 20 
xmin = 0  
xmax = 4

# Definition of the function domain
x_train=np.linspace(xmin,xmax,n_train)

# The real solution y(x) for training evaluation
y_train=tf.exp(-x_train)

# Input and output neurons (from the data)
input_neurons  = 1
output_neurons = 1

# Hiperparameters
epochs = 40

# Definition of the the model 
activation='elu'
input=Input(shape=(input_neurons,))
x=Dense(50, activation=activation)(input)
x=Dense(50, activation=activation)(x)
x=Dense(50, activation=activation)(x)
output = Dense(output_neurons,activation=None)(x)   
model=ODE_1st(input,output)

# Definition of the metrics, optimizer and loss
loss= tf.keras.losses.MeanSquaredError()
metrics=tf.keras.metrics.MeanSquaredError()    
optimizer= Adam(learning_rate=0.001)

model.compile(loss=loss,
          optimizer=optimizer, 
          metrics=[metrics])      
model.summary()

history=model.fit(x_train, y_train,batch_size=1,epochs=epochs)
\end{python}

\newpage 

The table~\ref{hyper1} summarized the hyperparameters used with this PINN.
\begin{table}[h]
\centering
\begin{tabular}{@{}ll}
\br
Parameter&Value\\
\mr
interval&(0,4)\\
n train&20\\
n hidden layers&3\\
n neurons/layer&1,50,50,50,1\\
epochs&50\\
activation&elu\\
optimizer&adam\\
learning rate&0,001\\
\br
\end{tabular}\\
\caption{\label{hyper1}Hyperparameters of the PINN to solve equation (\ref{ec:exp}).}
\end{table}

\begin{figure}[h]
  \centering
  \includegraphics[width=0.8\textwidth]{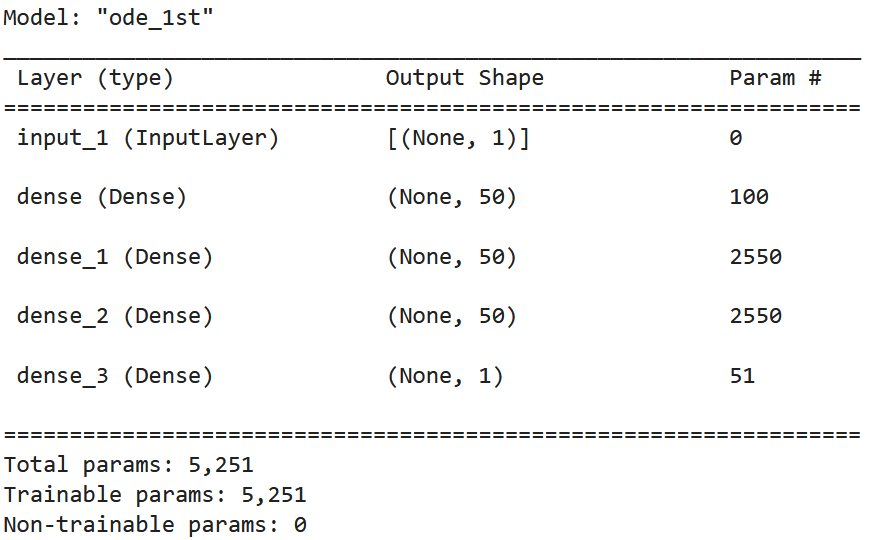}
  \caption{Keras-Tensorflow model of the PINN as given by the $model.summary()$ method of Keras.} 
  \label{model1}
\end{figure}

Figure~\ref{model1} provides an overview of the neural network architecture, which consists of 5251 trainable parameters. The shape of the outputs from each layer and their corresponding trainable parameters are also presented in the image. The image was generated with the $model.summary()$ method of Keras.

\newpage
\subsection{Evolution of losses during training} \label{sec:losses1}
Using the code below, the evolution of metrics as a function of the number of epochs can be obtained graphically.
\begin{python}
# summarize history for loss and metris
plt.rcParams['figure.dpi'] = 150
plt.plot(history.history['loss'],color='magenta',
         label='Total losses ($L_D + L_B$)')
plt.plot(history.history['mean_squared_error'],color='cyan',label='MSE')
plt.yscale("log")
plt.xlabel('epochs')
plt.legend(loc='upper right')
plt.show()
\end{python}
\begin{figure}[h!]
  \centering
  \includegraphics[width=0.75\textwidth]{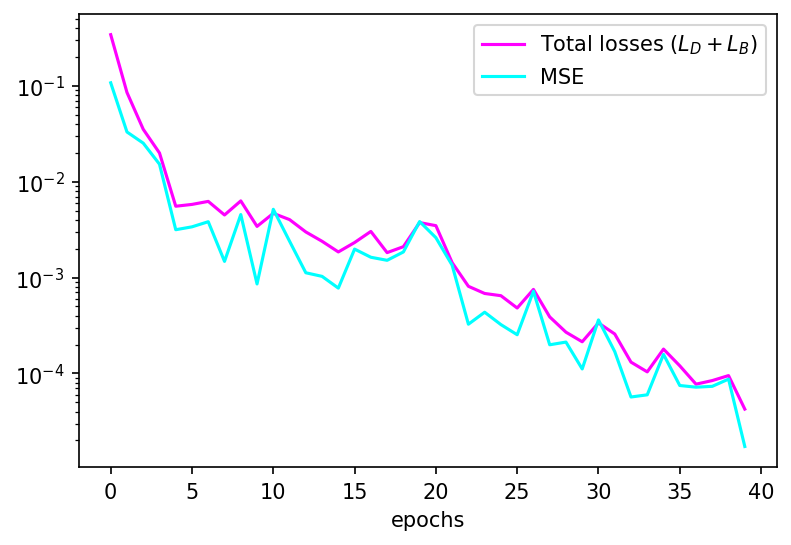}
  \caption{Training evolution as a function of the number of epochs. Two indicators of the performance of the NN are plotted: the total loss ($L_D$ + $L_B$) and the MSE. The last measures the distance per point between $y_{analytic}$ and $y_{NN}$, evaluated at the training points.} 
  \label{training1}
\end{figure}

\subsection{Solution and its derivatives} \label{sec:derivatives1}
TensorFlow allows the AD of any function $y(x)$, making it possible to calculate all of its derivatives. Since $y_{NN}(x)$ is simply a function, we can straightforwardly obtain them using this feature.

In the code, we define a set of validation points. Points not previously seen by the NN during training. The exact (analytical) values of $y(x)$ are also obtained. 

The two first derivatives are obtained with the help of two tf.GradientTape environments. These lines of the code look like the ones used in the Definition of the PINN (Section~\ref{defpinn1}).

\begin{python}
# Check the PINN at different points not included in the training set
n = 500
x=np.linspace(0,4,n)
y_exact=tf.exp(-x)
y_NN=model.predict(x)

# The gradients (y'(x) and y''(x)) from the model 
x_tf = tf.convert_to_tensor(x, dtype=tf.float32)
with tf.GradientTape(persistent=True) as t:  
  t.watch(x_tf)  
  with tf.GradientTape(persistent=True) as t2:
        t2.watch(x_tf)
        y = model(x_tf)
  dy_dx_NN = t2.gradient(y, x_tf)
d2y_dx2_NN = t.gradient(dy_dx_NN, x_tf)

# Plot the results
plt.rcParams['figure.dpi'] = 150
plt.plot(x, y_exact, color="black",linestyle='solid', 
                     linewidth=2.5,label="$y(x)$ analytical")
plt.plot(x, y_NN, color="red",linestyle='dashed',
                     linewidth=2.5, label="$y_{NN}(x)$")  
plt.plot(x, dy_dx_NN, color="blue",linestyle='-.',
                     linewidth=3.0, label="$y'_{NN}(x)$")
plt.plot(x, d2y_dx2_NN, color="green", linestyle='dotted',
                     linewidth=3.0, label="$y''_{NN}(x)$")
plt.legend()
plt.xlabel("x")
plt.show()  
\end{python}

\begin{figure}[h!]
  \centering
  \includegraphics[width=0.75\textwidth]{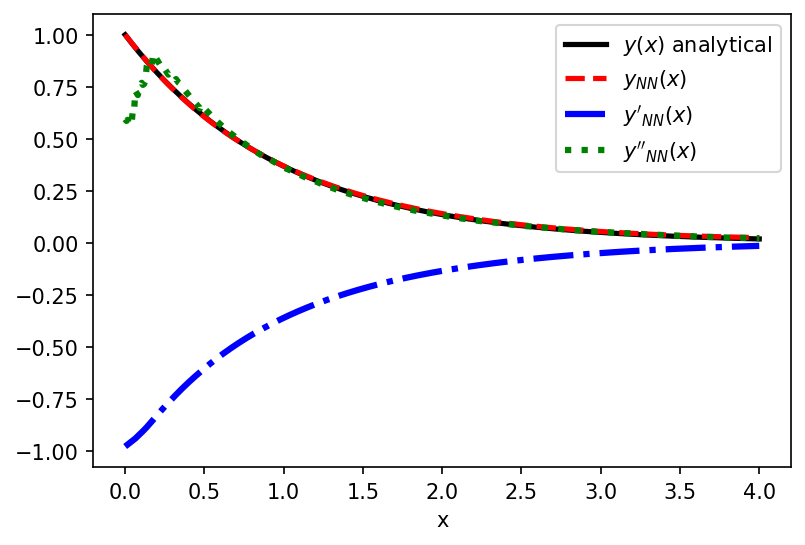}
  \caption{The analytical solution and the prediction for $y(x)$ and its derivatives as a function of $x$. The results are obtained from previously unseen points inside and outside the training interval, which were utilized as validation data for the PINN.} 
  \label{y1}
\end{figure}

\section{Example 2: 2nd order linear Ordinary Differential Equations}
Here we solve the second example described in the paper, the solution to the harmonic oscillator problem using PINNs:
\begin{equation}\label{ec:osc}
    \left\{
        \begin{array}{l}
            y''(x) + y(x)=0 \hspace{0.5cm}with\hspace{0.5cm} 0\leqslant x \leqslant 8 \\
            y(0)=1, y'(0)=0
        \end{array} 
        \right.\qquad
\end{equation}
, whose analytic solution is $y(x)=\cos(x)$.

The codes for the 2nd-order ODE examples are analog to the example of the previous section. However, we need to pay special attention to the definition of the total loss.

\subsection{Main libraries}
We use here the same list of libraries that in the first example.
\begin{python}
# Tensorflow Keras and the rest of the packages
import tensorflow as tf
from tensorflow.keras.layers import Dense,Input
from tensorflow.keras.optimizers import Adam
import numpy as np
import matplotlib.pyplot as plt
\end{python}

\subsection{Definition of the PINN}
As in the previous case, this is the most technical step in the code where the loss is defined, so it incorporates the differential equation.

We have introduced several significant changes to our approach. Firstly, we utilized an extra \textit{tf.GradientTape} to calculate the second-order derivative required to define the loss. Additionally, we incorporated a new term to account for the initial velocity condition, and we made modifications to the differential equation term. Specifically, we are now solving for $y''+y=0$, and as a result, we are utilizing $self.compiled_{-}loss(d2y_{-}dx2, -y)$ in our computations.
\begin{python}
class ODE_2nd(tf.keras.Model):                  
    def train_step(self, data):
        # Training points and the analytical 
        # (exact) solution at this points
        x, y_exact = data 
        #Change initial conditions for the PINN       
        x0=tf.constant([0.0], dtype=tf.float32) 
        y0_exact=tf.constant([1.0], dtype=tf.float32) 
        dy_dx0_exact=tf.constant([0.0], dtype=tf.float32)   
        # Calculate the gradients and update weights and bias      
        with tf.GradientTape() as tape:   
            tape.watch(x)
            tape.watch(y_exact)
            tape.watch(x0)
            tape.watch(y0_exact)
            tape.watch(dy_dx0_exact)
            # Calculate the gradients dy2/dx2, dy/dx    
            with tf.GradientTape() as tape0:
                    tape0.watch(x0)         
                    y0_NN = self(x0,training=False)          
                    tape0.watch(y0_NN)
            dy_dx0_NN = tape0.gradient(y0_NN, x0)       
            with tf.GradientTape() as tape1:    
                tape1.watch(x)                            
                with tf.GradientTape() as tape2:
                    tape2.watch(x)         
                    y_NN = self(x,training=False)          
                    tape2.watch(y_NN)
                dy_dx_NN = tape2.gradient(y_NN, x)                 
                tape1.watch(y_NN)
                tape1.watch(dy_dx_NN)                
            d2y_dx2_NN = tape1.gradient(dy_dx_NN, x)                                    
            tape.watch(y_NN)
            tape.watch(dy_dx_NN)  
            tape.watch(d2y_dx2_NN)  

            #Loss= ODE+ boundary/initial conditions                     
            y0_exact=tf.reshape(y0_exact,shape=y_NN[0].shape)            
            dy_dx0_exact=tf.reshape(dy_dx0_exact,shape=dy_dx0_NN.shape)        
          
            loss= self.compiled_loss(y0_NN,y0_exact)\
                  +self.compiled_loss(d2y_dx2_NN,-y_NN)\
                  +self.compiled_loss(dy_dx0_NN,dy_dx0_exact)
                  
        gradients = tape.gradient(loss, self.trainable_weights)
        self.optimizer.apply_gradients(zip(gradients, self.trainable_weights))
        self.compiled_metrics.update_state(y_exact, y_NN)
        return {m.name: m.result() for m in self.metrics}
\end{python}

\subsection{Run the PINN}
We used in solving the 2nd-order ODE an specific~\textit{callback} of Keras, which stops the training if there is no improvement and saves the NN's configuration with the lowest loss.

Here, we also introduced a different initializer of the weights and biases called GlorotUniform. There are many initializers, and part of improving the NN predictions consists of finding the best initializer in each case. 

\begin{python}
n_train = 18
xmin = 0.0  
xmax = 8.0

# Definition of the function domain
x_train=np.linspace(xmin,xmax,n_train)

# The solution y(x) for training and validation datasets
y_train=np.cos(x_train)

# Input and output neurons (from the data)
input_neurons  = 1
output_neurons = 1

# Hiperparameters
epochs = 500

# Definition of the the model 
initializer = tf.keras.initializers.GlorotUniform(seed=5)
activation='tanh'
input=Input(shape=(input_neurons,))
x=Dense(50, activation=activation,
            kernel_initializer=initializer)(input)
x=Dense(50, activation=activation,
            kernel_initializer=initializer)(x)
x=Dense(50, activation=activation,
            kernel_initializer=initializer)(x)
output = Dense(output_neurons,
               activation=None,
               kernel_initializer=initializer)(x)    
model=ODE_2nd(input,output)

# Definition of the metrics, optimizer and loss
loss= tf.keras.losses.MeanSquaredError()
metrics=tf.keras.metrics.MeanSquaredError()    
optimizer= Adam(learning_rate=0.001)

model.compile(loss=loss,
          optimizer=optimizer, 
          metrics=[metrics])      
model.summary()

# Just use `fit` as usual
callback = tf.keras.callbacks.EarlyStopping(monitor='loss', 
                                            patience=1000,
                                            restore_best_weights=True)

history=model.fit(x_train, y_train,batch_size=1, epochs=epochs,
                  callbacks=callback)
\end{python}

The hyperparameters for the 2nd-order ODE are summarized in Table~\ref{hyper2}, with slight modifications compared to the previous example.

\begin{table}[h]
\centering
\begin{tabular}{@{}ll}
\br
Parameter&Value\\
\mr
interval&(0,8)\\
n train&18\\
n hidden layers&3\\
n neurons/layer&1,50,50,50,1\\
epochs&500\\
activation&tanh\\
optimizer&adam\\
learning rate&0,001\\
\br
\end{tabular}\\
\caption{\label{hyper2}Hyperparameters of the PINN to solve equation (\ref{ec:osc}).}
\end{table}

\subsection{Evolution of losses during training}
A comparable evolution, similar to the one depicted in the article in Fig.~3, can be observed by utilizing the same code as presented in Section~\ref{sec:losses1}.
\begin{figure}[h!]
  \centering
  \includegraphics[width=0.7\textwidth]{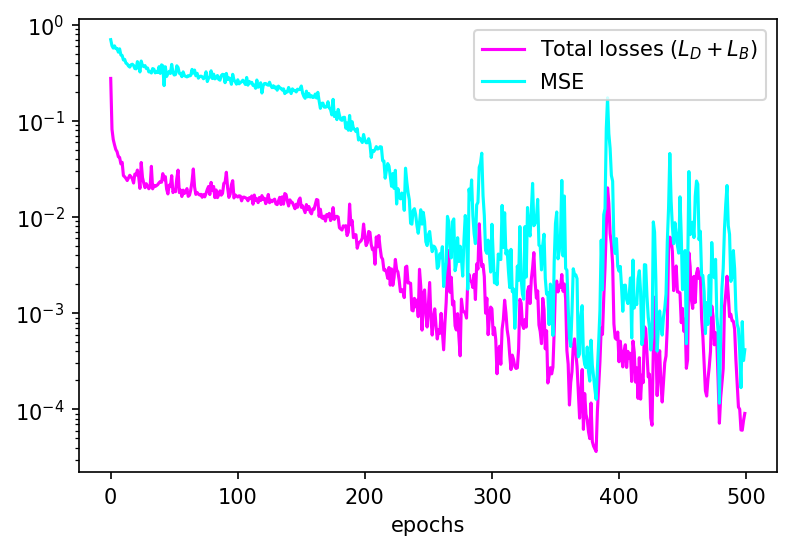}
  \caption{Evolution of losses during training, as a function of the number of epochs.} 
  \label{y2}
\end{figure}

\newpage
\subsection{Solution and its derivatives}
Using the code of Section~\ref{sec:derivatives1}, results like those shown in the paper in Fig.~3 can be obtained. It is important to note that the exact solution for the 2n-order ODE must be employed in this case, which is:

\begin{python}
y_exact=tf.cos(-x)
\end{python}

\begin{figure}[h!] \label{y2}
  \centering
  \includegraphics[width=0.7\textwidth]{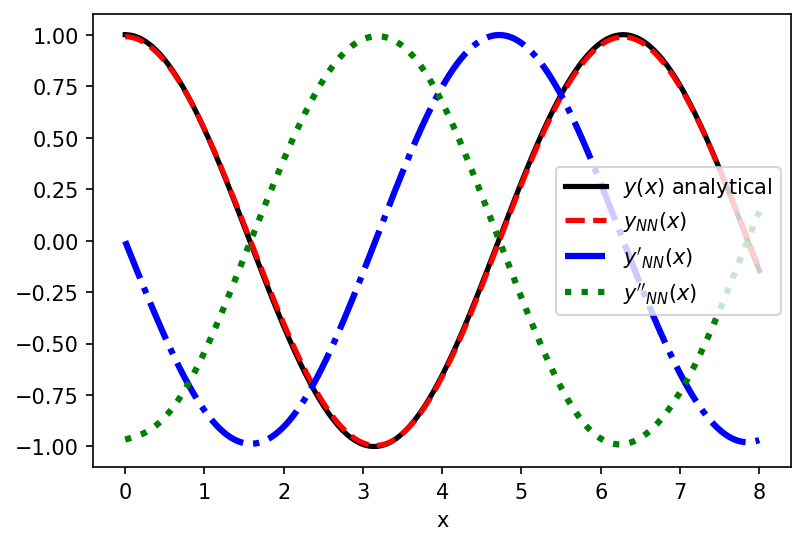}
  \caption{The analytical solution and the prediction for $y(x)$ and its derivatives as a function of $x$.} 
  \label{der2}
\end{figure}

\newpage
\section{Example 3: 2nd order non-linear Ordinary Differential Equations}
Finally, we describe the Python code to solve the  2nd-order nonlinear ODE used as an example in the article.
\begin{equation}\label{ec:gp}
    \left\{
        \begin{array}{l}
            y''(x) - y(x) - 3y^2(x)=0  \\            
            \hspace{1cm}y(0)=-1/2, y'(0)=0
        \end{array} 
        \right.\qquad
\end{equation}
, being its analytical solution $y(x) = - \frac{1}{2}\cdot sech ^2(\frac{x}{2})$.

\subsection{Main libraries}
These are the same packages as those used in the previous examples.

\begin{python}
# Tensorflow Keras and rest of the packages
import tensorflow as tf
from tensorflow.keras.layers import Dense,Input
from tensorflow.keras.optimizers import Adam
import numpy as np
import matplotlib.pyplot as plt
\end{python}

\subsection{Definition of the PINN}
As in the previous case, this is the most technical step in the code where the loss is defined, so it incorporates the differential equation.
\begin{python}
class ODE_2nd(tf.keras.Model):                  
    def train_step(self, data):
        # Training points and the analytical 
        # (exact) solution at this points
        x, y_exact = data                
        #Change initial conditions for the PINN
        x0=tf.constant([0.0], dtype=tf.float32) 
        y0_exact=tf.constant([-0.5], dtype=tf.float32) 
        dy_dx0_exact=tf.constant([0.0], dtype=tf.float32) 
        C=tf.constant([1.0], dtype=tf.float32)
        # Calculate the gradients and update weights and bias
        with tf.GradientTape() as tape:   
            tape.watch(x)
            tape.watch(y_exact)
            tape.watch(x0)
            tape.watch(y0_exact)
            tape.watch(dy_dx0_exact)
            # Calculate the gradients dy2/dx2, dy/dx    
            with tf.GradientTape() as tape0:
                    tape0.watch(x0)         
                    y0_NN = self(x0,training=False)          
                    tape0.watch(y0_NN)
            dy_dx0_NN = tape0.gradient(y0_NN, x0)       
            with tf.GradientTape() as tape1:    
                tape1.watch(x)                            
                with tf.GradientTape() as tape2:
                    tape2.watch(x)         
                    y_NN = self(x,training=False)          
                    tape2.watch(y_NN)
                dy_dx_NN = tape2.gradient(y_NN, x)                 
                tape1.watch(y_NN)
                tape1.watch(dy_dx_NN)                
            d2y_dx2_NN = tape1.gradient(dy_dx_NN, x)                                    
            tape.watch(y_NN)
            tape.watch(dy_dx_NN)  
            tape.watch(d2y_dx2_NN)                     

            #Loss= ODE+ boundary/initial conditions                      
            y0_exact=tf.reshape(y0_exact,shape=y_NN[0].shape)            
            dy_dx0_exact=tf.reshape(dy_dx0_exact,shape=dy_dx0_NN.shape)
            C=tf.reshape(C,shape=d2y_dx2_NN.shape)

            loss= self.compiled_loss(y0_NN,y0_exact)\
                  +self.compiled_loss(dy_dx0_NN,dy_dx0_exact)\
                  +self.compiled_loss(d2y_dx2_NN,C*y_NN+3.0*y_NN**2) 

        gradients = tape.gradient(loss, self.trainable_weights)
        self.optimizer.apply_gradients(zip(gradients, self.trainable_weights))
        self.compiled_metrics.update_state(y_exact, y_NN)
        return {m.name: m.result() for m in self.metrics}           
\end{python}

\subsection{Run the PINN}
Regarding the solution found for the 2nd-order ODE, the number of epochs increases because the learning rate is reduced. 
This example, however, is not so sensitive to the initial values of weights and biases, so we removed the GlorotUniform initializer. 

\begin{python}
n_train = 11 
xmin = -5  
xmax = 5

# Definition of the function domain
x_train=np.linspace(xmin,xmax,n_train)

# The solution y(x) for training and validation datasets
x0=0.0
y_train=-0.5*1.0*(1.0/np.cosh(0.5*np.sqrt(1.0)*(x_train-x0)))**2

# Input and output neurons (from the data)
input_neurons  = 1
output_neurons = 1

# Hiperparameters
epochs = 2000

# Definition of the model 
activation='tanh'
input=Input(shape=(input_neurons,))
x=Dense(50, activation=activation)(input)
x=Dense(50, activation=activation)(x)
x=Dense(50, activation=activation)(x)
output = Dense(output_neurons,activation=None)(x)    
model=ODE_2nd(input,output)

# Definition of the metrics, optimizer and loss
loss=tf.keras.losses.MeanSquaredError()
metrics=tf.keras.metrics.MeanSquaredError()    
optimizer= Adam(learning_rate=0.0001) 

model.compile(loss=loss,
          optimizer=optimizer, 
          metrics=[metrics])      
model.summary()

# Just use `fit` as usual
callback = tf.keras.callbacks.EarlyStopping(monitor='loss', 
                                            patience=1000,
                                            restore_best_weights=True)

history=model.fit(x_train, y_train, batch_size=1, epochs=epochs,
                  callbacks=callback)
\end{python}

Table \ref{hyper3} sums up the hyperparameters for this ODE.
\begin{table}[h!]
\centering
\begin{tabular}{@{}ll}
\br
Parameter&Value\\
\mr
interval&(-5,5)\\
n train&11\\
n hidden layers&3\\
n neurons/layer&1,50,50,50,1\\
epochs&2000\\
activation&tanh\\
optimizer&adam\\
learning rate&0,0001\\
\br
\end{tabular}\\
\caption{\label{hyper3}Hyperparameters of the PINN to solve equation (\ref{ec:gp}).}
\end{table}

\subsection{Evolution of losses during training}
A comparable evolution, similar to the one depicted in the article in Fig.~3, can be observed by utilizing the same code as presented in Section~\ref{sec:losses1}.
\begin{figure}[h!]
  \centering
  \includegraphics[width=0.7\textwidth]{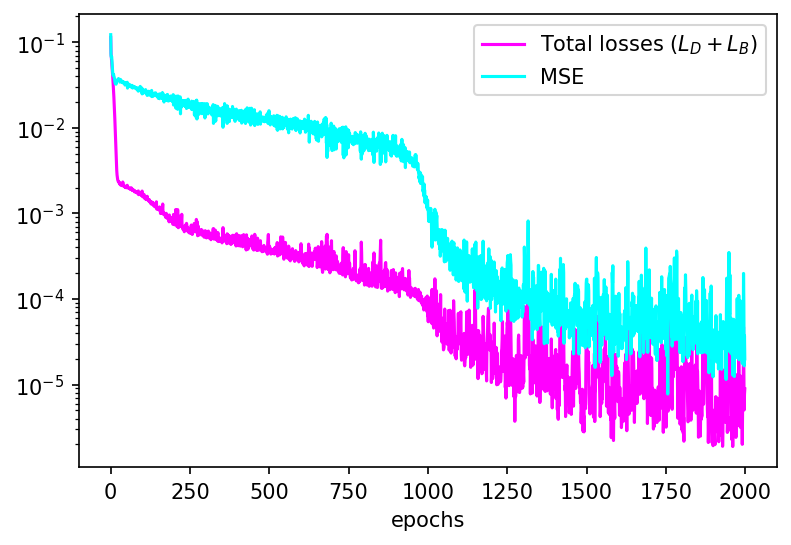}
  \caption{Training evolution as a function of the number of epochs.} 
  \label{training3}
\end{figure}
\newpage
\subsection{Solution and its derivatives}
To obtain comparable results to those presented in Fig.~3 of the article, one can use the code from Section\ref{sec:derivatives1}. However, it is necessary to include the exact solution of the 2nd-order nonlinear ODE:
\begin{python}
x0=0.0
y=-0.5*1.0*(1.0/np.cosh(0.5*np.sqrt(1.0)*(x-x0)))**2
\end{python}
\begin{figure}[h!]
  \centering
  \includegraphics[width=0.7\textwidth]{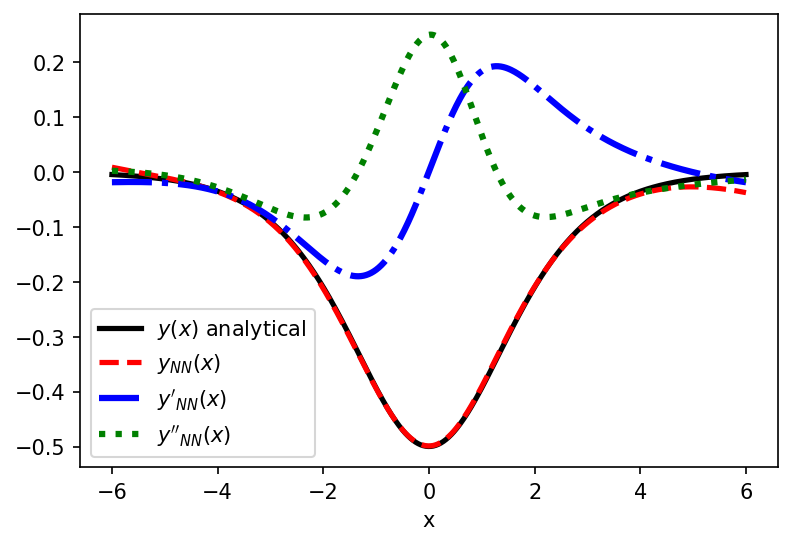}
  \caption{The analytical solution and the prediction for $y(x)$ and its derivatives as a function of $x$.} 
  \label{y3}
\end{figure}

\section{Performance of the PINNs}
 We conducted 100 simulations for each example, enforcing a total loss threshold of $1.0\cdot 10^{-4}$ for Examples 1 and 2 and $1.0\cdot 10^{-5}$ for Example 3.

\begin{figure}[h!] 
  \centering
  \includegraphics[width=1.0\textwidth]{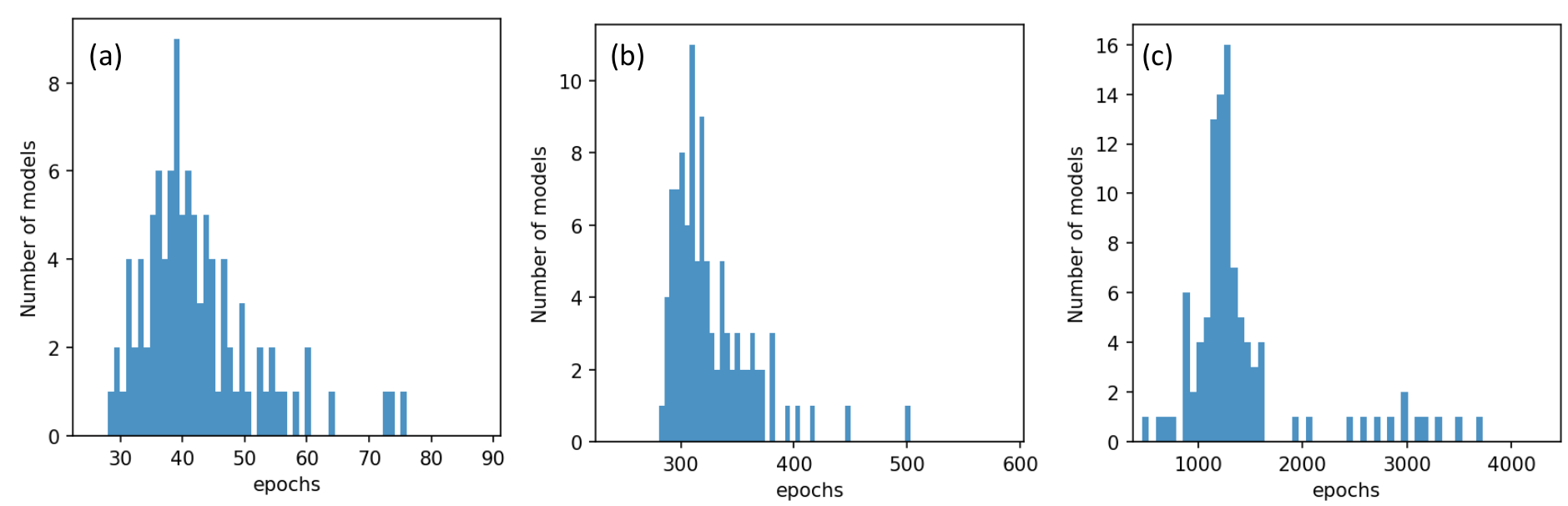}
  \caption{The histograms display the distribution of the number of trained models that achieved the convergence threshold within a specific number of epochs. This data was obtained for: (a) the 1st-order ODE studied, corresponding to the exponential decay solution; (b) the 2nd-order ODE (harmonic oscillator); and (c) the 2nd-order nonlinear ODE (Korteweg-de Vries equation).} 
  \label{histogram}
\end{figure}

The numerical experiments are presented in Figure~\ref{histogram}, which indicate that, on average, the threshold is reached after 42, 326, and 1422 epochs. However, it's worth noting that the distributions are non-normal, and a more insightful analysis requires the use of median and mode statistics, where the median represents the middle value of the dataset. The mode defines the most common value. For instance, the median and mode for the Example 1 are 40 and 39 epochs. For the Example 2 we have  317 and 312. Finally, for the Example 3, the median and mode are 1256 and 1225 respectively.

\newpage
\section{References}